\documentclass[aps,prl,twocolumn,showpacs,groupedaddress]{revtex4}  
\usepackage{graphicx}  
\usepackage{dcolumn}   
\usepackage{bm}        
\usepackage{amsmath}   
\usepackage{amssymb}   
\usepackage{rotating}
\newcommand{\MET} {\mbox{$\not \!\! E_T$}}

\newcommand{\ppbar} {p\bar{p}}
\newcommand{\ttbar} {t\bar{t}}

\def        \dzero  {D0~}

\newcommand{\ifm}[1]{\relax\ifmmode #1\else $#1$\hskip 0.15cm\fi}

\newcommand{\tbp}{$\tan\beta$}
\newcommand{\mH}{\ifm{m_{H^{\pm}}}}

\newcommand{\mHp}{$m_{H^{\pm}}$}

\newcommand{\mt}{\ifm{m_t}}

\newcommand{\mb}{\ifm{m_b}}

\newcommand{\tHb}{\ifm{t \rightarrow  H^+b}}

\newcommand{\Htn}{\ifm{H^{+} \rightarrow \tau^{+} \nu_\tau}}

\newcommand{\beq}{\begin{equation}}
\newcommand{\ee}{\end{equation*}}
\newcommand{\eeq}{\end{equation}}
\newcommand{\bdm}{\begin{displaymath}}
\newcommand{\edm}{\end{displaymath}}

\newcommand{\fracerr}[1]{\ifmmode
                    \frac{\delta #1}{#1}
               \else
                    \mbox{${\delta #1}/{#1}$}
               \fi}

\newcommand{\mpt}{\slash\kern-5pt p_T}

\newcommand{\lsim}{\mathrel{\hbox{\rlap{\lower.55ex\hbox{$\sim$}} \kern-.3em 
\raise.4ex \hbox{$<$}}}}

\begin{document}
%


\hspace{5.2in} \mbox{FERMILAB-PUB-09/329-E}

\title{Search for charged Higgs bosons in decays of top quarks}

%
\author{V.M.~Abazov$^{37}$}
\author{B.~Abbott$^{75}$}
\author{M.~Abolins$^{65}$}
\author{B.S.~Acharya$^{30}$}
\author{M.~Adams$^{51}$}
\author{T.~Adams$^{49}$}
\author{E.~Aguilo$^{6}$}
\author{M.~Ahsan$^{59}$}
\author{G.D.~Alexeev$^{37}$}
\author{G.~Alkhazov$^{41}$}
% ...

\affiliation{\vspace{0.1 in}(The D\O\ Collaboration)\vspace{0.1 in}}
% ...
\affiliation{$^{6}$University of Alberta, Edmonton, Alberta, Canada;
                Simon Fraser University, Burnaby, British Columbia, Canada;
                York University, Toronto, Ontario, Canada and
                McGill University, Montreal, Quebec, Canada}
% ...
\affiliation{$^{30}$Tata Institute of Fundamental Research, Mumbai, India}
% ...
\affiliation{$^{37}$Joint Institute for Nuclear Research, Dubna, Russia}
% ...
\affiliation{$^{41}$Petersburg Nuclear Physics Institute,
                St. Petersburg, Russia}
% ...
\affiliation{$^{49}$Florida State University, Tallahassee, Florida 32306, USA}
% ...
\affiliation{$^{51}$University of Illinois at Chicago,
                Chicago, Illinois 60607, USA}
% ...
\affiliation{$^{59}$Kansas State University, Manhattan, Kansas 66506, USA}
% ...
\affiliation{$^{65}$Michigan State University,
                East Lansing, Michigan 48824, USA}
% ...
\affiliation{$^{75}$University of Oklahoma, Norman, Oklahoma 73019, USA}
% ...

\date{30 June 2009}

\begin{abstract}
We present a search for charged Higgs bosons in decays of top quarks, 
in the mass range $80 < \mH < 155$~GeV, assuming the subsequent decay 
$\Htn$ (and its charge conjugate). Using 0.9~fb$^{-1}$ of lepton+jets 
data collected with the D0 detector at the Fermilab Tevatron $\ppbar$ 
collider, operating at a center of mass energy $\sqrt{s}=1.96$~TeV, 
we find no evidence for a $H^{\pm}$ signal. Hence we exclude branching 
ratios $B(\tHb) > 0.24$ for $\mH=80$~GeV and $B(\tHb) > 0.19$ for 
$\mH=155$~GeV at the $95\%$ C.L.
\end{abstract}

\maketitle

\clearpage
\normalsize

The electroweak symmetry breaking sector of the standard model (SM)
contains a single SU(2) complex scalar doublet field that provides 
gauge-invariant generation of particle masses, with the only 
observable particle being the electrically neutral Higgs boson 
$H^0$~\cite{higgs}. Here, we search for evidence of a richer structure.
The simplest
extension to the SM Higgs sector involves the addition of a second SU(2)
complex scalar doublet, which introduces five spin-0 particles, three
that are neutral and two that are charged ($H^{\pm}$)~\cite{thdm}.
The fermion couplings to the Higgs doublets are not specified {\it a priori}, 
and the only requirement is that flavor changing neutral currents 
are not allowed at lowest level in perturbation theory. 
One possibility, the Type-II model, couples
the up-type fermions to one Higgs doublet and the down-type
fermions to the other, as required in the minimal supersymmetric
extension to the SM (MSSM)~\cite{thdm}. 
In addition, the MSSM constrains the five Higgs masses through
two free parameters:
$\tan{\beta}$, the ratio of the vacuum expectation values of the 
two doublets, and the mass of any one of the physical Higgs 
bosons. We choose \mH for the latter.

Since the Yukawa coupling to the $H^{\pm}$ boson increases with 
fermion mass for all values of \tbp, top 
and bottom quarks in this model are expected to have large
Yukawa couplings. Therefore, if 
$\mH<\mt - \mb$, the decay $\tHb$ (and its charge conjugate) is expected to 
have a large branching fraction for all \tbp. 
Further, for large values of \tbp~(\tbp~$\geq 10$), the charged Higgs decays 
predominantly to a $\tau$ lepton and its associated neutrino with 
$B(\Htn) \approx 1$. Hence if the $H^{\pm}$ boson exists and $B(\tHb)$ 
is substantial, a search optimized 
for the study of SM decays of $\ttbar$ to lepton+jets final states should  
show a deficit of events relative to the SM prediction 
because of differences in decay branching fractions 
and in kinematic distributions. Any such deficit 
could therefore be indicative of the presence of charged Higgs 
bosons in decays of the top quark.

Direct searches for $H^{\pm}$ bosons have been performed
at the LEP $e^+e^-$ collider at CERN~\cite{LEP} and at the Tevatron 
$\ppbar$ collider at Fermilab~\cite{fermilab}.
With no evidence of a signal, the LEP experiments set a combined 
limit of $\mH>78.6$~GeV independent of $B(\Htn)$,
while the Tevatron experiments have set 
limits in the context of a Type-II two Higgs doublet model 
that exclude regions of the [\tbp , $\mH$] parameter space~\cite{LEPLimit}.
Searches for indirect evidence of $H^{\pm}$ bosons through radiative 
decays of $B$ mesons at $B$ factories provide a combined limit of 
$\mH>295$~GeV~\cite{Belle,BaBar,BLimit}. Although
$B$ factories exclude a larger
part of parameter space than our current
study, it is important to search for objects
such as the $H^\pm$ bosons through all possible channels
and not defer entirely to theory.

\begin{figure} [htb]
\vspace{-0.7in}
\begin{center}
\includegraphics[width=0.45\textwidth]  
{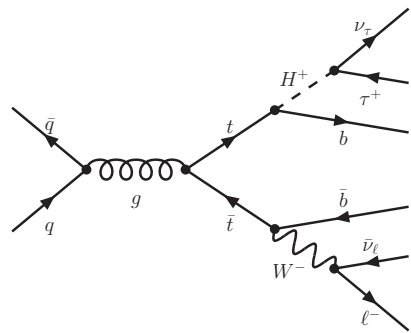}
\end{center}
\vspace*{-1.7in}
\caption{Representative Feynman diagram for charged Higgs boson 
production in top quark decays at the Tevatron ($\ell = e$ or $\mu$).}
\label{fig:feyn}
\end{figure}

In this article, we describe the search for charged Higgs bosons 
from top quark decays in $\ttbar$ events with one lepton 
(electron $e$ or muon $\mu$) and at least three jets. 
A representative Feynman diagram for such events is shown in 
Fig.~\ref{fig:feyn}, where one of the top quarks decays to a 
$W$ boson and a $b$ quark, as in the SM, 
and the other decays to a $H^{\pm}$ boson 
and a $b$ quark. For our signal, we consider events in which the 
$W$ boson decays leptonically 
($e$, $\mu$, or $\tau$, with the $\tau$ decaying to 
an $e$ or $\mu$ and two neutrinos), while the charged Higgs 
boson decays to a $\tau$ and a neutrino and
the $\tau$ decays to a neutrino and hadrons. The final state
therefore consists of an isolated lepton ($e$ or $\mu$) with 
large transverse momentum ($p_T$), significant missing 
transverse energy ($\MET$) from the escaping neutrinos, and at least 
three jets: two from the $b$ quarks and one from the decay of the $\tau$. 
No attempt is made to identify $\tau$ leptons in such decays.
Some of the signal can also come from events where the $\tau$ 
from the $H^\pm$ boson decays leptonically, while the $W$ boson decays into
a quark-antiquark pair, thereby giving two jets.
In that case, there will be four jets in the final state.
Finally, if both top quarks decay into charged Higgs bosons, which 
then decay into $\tau$ leptons, and one $\tau$ decays leptonically 
while the other decays into a jet, this can also contribute to 
the signal.
The largest backgrounds to these processes are from SM decays 
of $\ttbar$ pairs and $W$+jets production, along with smaller 
contributions from the production of single top quarks, dibosons 
($WW$, $WZ$, and $ZZ$), and $Z$+jets. 
An additional source of background is from multijet events, in 
which a jet mimics an electron, or a muon from $b$ (or $c$) quark 
decay appears to be isolated. 

We analyze $0.90 \pm 0.05 ~\rm fb^{-1}$ of data recorded with 
the D0 detector~\cite{Andeen:2006,D0detector}. 
The trigger required a 
reconstructed jet and an electromagnetic energy cluster in the 
electron channel or a jet and a muon candidate in the muon channel.  
We base this analysis on a previous one that extracted the 
$\ttbar$ production cross section within the
framework of the SM, i.e., assuming 
$B(t\to W^+b)=1$~\cite{topologicalXsec}.
The principal difference is that here we consider an additional 
decay mode ($t\to H^+b$) and attempt to measure 
$B \equiv B(t\to H^+b)$ under the constraint 
$B(t\to W^+b)+B(t\to H^+b)=1$.
For any measurement of $B$, $m_{H^{\pm}}$ 
is treated as a fixed parameter. Measurements are made for several 
values of $m_{H^{\pm}}$. 

We apply the same event selection criteria as in 
Ref.~\cite{topologicalXsec} to separate $\ttbar$ production from 
background. 
These are summarized in Table~\ref{tab:selections}. We impose an 
additional requirement of $\sum p_T{\rm (jet)} > 120~{\rm GeV}$ 
for events with only three jets and separate the events into 
two jet-multiplicity bins (3 jets and $>3$ jets) to improve 
signal discrimination. 
\begin{table}[!h!tbp]  
\begin{center}  
\caption{Summary of event selections.}   
\begin{tabular}{l|cc}\hline\hline 
&$e$ + jets channel & $\mu$ + jets channel \\ \hline 
Lepton ($\ell$) &$p_T>$ 20 GeV & $p_T>$ 20 GeV   \\ 
&$|\eta|<$ 1.1~\cite{d0system} & $|\eta|<$ 2.0 \\\hline 
$\MET$ & $\MET >$ 20 GeV & $\MET >$ 25 GeV \\ 
$\Delta\phi(\ell, \MET )$~\cite{d0system} 
& $> 0.7\pi - 0.045\MET$ 
&  \hskip 0.5cm $> 2.1\pi - 0.033\MET$ \\
($\MET$ in GeV) & &   \\ \hline 
Jets& \multicolumn{2}{c}{$>$ 2, $p_T>$ 20 GeV, $|\eta|<$ 2.5}\\ 
&\multicolumn{2}{c}{$p_T$(jet1) $>$ 40 GeV}\\ 
\hline\hline 
\end{tabular} 
\label{tab:selections} 
\end{center} 
\vspace*{-0.5cm} 
\end{table} 

To model the background distributions, $W$+jets and $Z$+jets 
events are generated using {\sc alpgen}~\cite{ALPGEN-MLM}, 
while {\sc singletop}~\cite{SINGLETOP} is used for single top quark 
events. 
The events are passed through {\sc pythia}~\cite{PYTHIA} for parton 
showering and hadronization. 
Diboson and SM $\ttbar$ events are generated using {\sc pythia}. 
The non-SM decay modes of $\ttbar$ events where one or both 
top quarks decay to a charged Higgs boson, are also generated using {\sc pythia}.  
The Monte Carlo (MC) events for the $H^{\pm}$ signal are produced at 
the following values of $m_{H^{\pm}}$: 80, 100, 120, 140, 150, and 155~GeV. 
All MC generated events are processed through the D0 
detector simulation based on {\sc geant}~\cite{GEANT}, followed by 
application of the same reconstruction algorithms as used on D0 
data. Subsequent corrections are also applied to MC events to 
account for trigger efficiencies and differences between 
MC events and data in object reconstruction efficiencies and resolutions.

To determine the number of background multijet events, we use a data sample 
with looser electron identification or weaker muon isolation criteria, 
as described in Ref.~\cite{topologicalXsec}. The normalization 
of the $W$+jets contribution is determined differently in the current 
analysis, as discussed below. 
For the prediction of yields for the single top quark, diboson, and 
$Z$+jets events, we use 
next-to-leading order cross sections~\cite{bkg_xsec}. The number of 
$\ttbar$ events is obtained by summing the different top quark decay 
modes according to their accepted branching fractions and 
respective selection efficiencies ($\epsilon$) as follows:
%
\begin{multline}
  N_{t\bar t} = \left[ (1-B)^2\cdot\epsilon_{WW}   
  + 2(1-B)B\cdot\epsilon_{WH}\right.
  \\\left.+ B^2\cdot\epsilon_{HH}\right] 
  \cdot\sigma(\ttbar)\cdot\int\mathcal{L}dt,  
  \label{eq:signalttbar}  
\end{multline}
%
where {\it WW} represents SM decays of the 
top quark, {\it WH} and {\it HH} represent non-SM decays of one 
or both top quarks, respectively, and $\int\mathcal{L}dt$ is the 
integrated luminosity. 
We use $\sigma(\ttbar) = 7.48^{+0.55}_{-0.72}~\rm pb$ for 
a top quark mass of $m_t~=~172.4 \pm 1.2~{\rm GeV}$~\cite{ttbar-theory}, and 
consider $B$ as the parameter of interest for any fixed value 
of $m_{H^{\pm}}$. The selection efficiencies for the {\it WW} 
decay modes in the different channels are $\approx 2\%$, 
which includes all corrections and trigger effects. 
The corresponding efficiencies for the {\it WH} ({\it HH}) 
modes  vary between 1.5\%--0.5\% (1.2\%--0.3\%) 
for different values of $m_{H^{\pm}}$.

\begin{figure} [htb]
\includegraphics[width=0.40\textwidth]  
{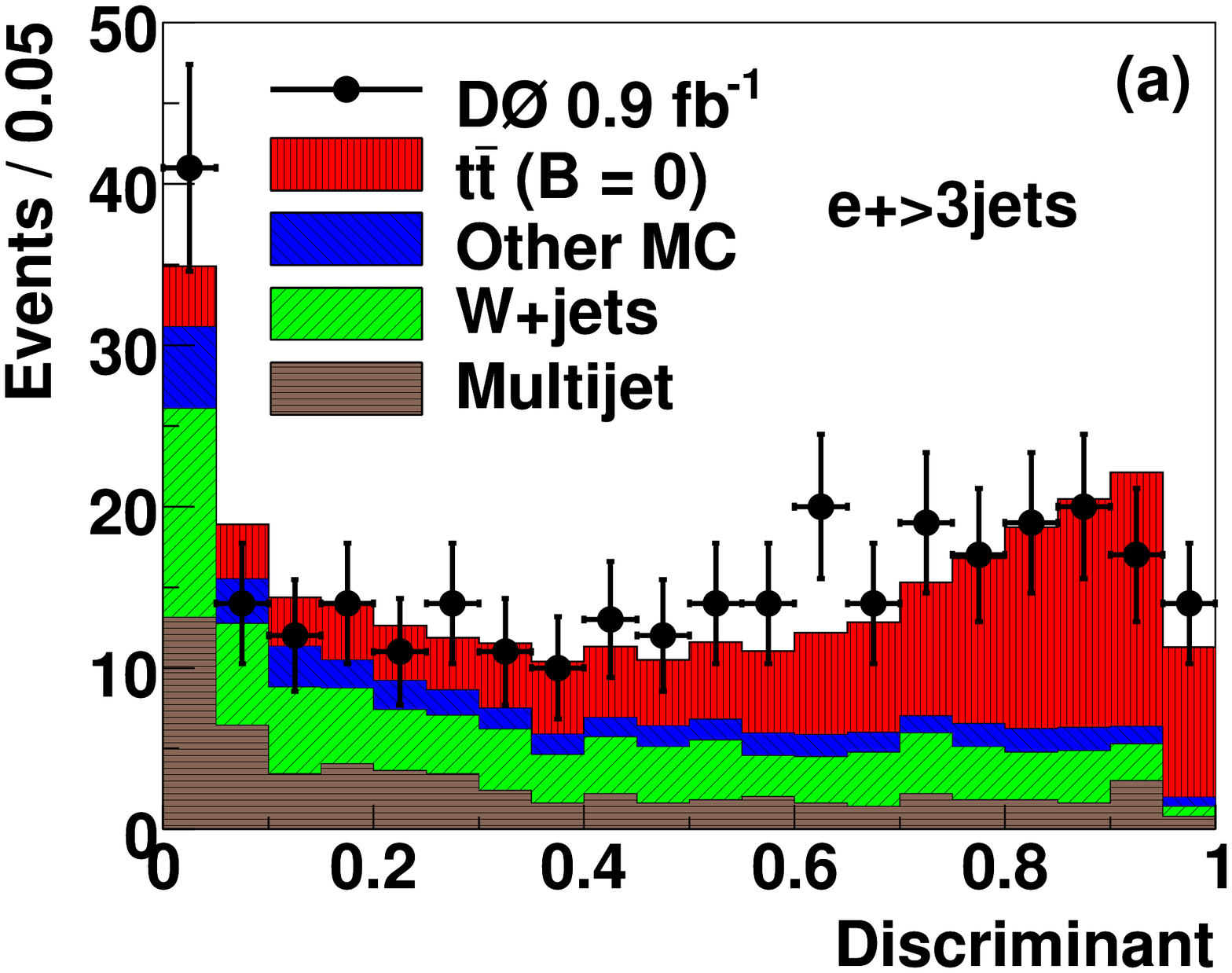}
\includegraphics[width=0.40\textwidth]  
{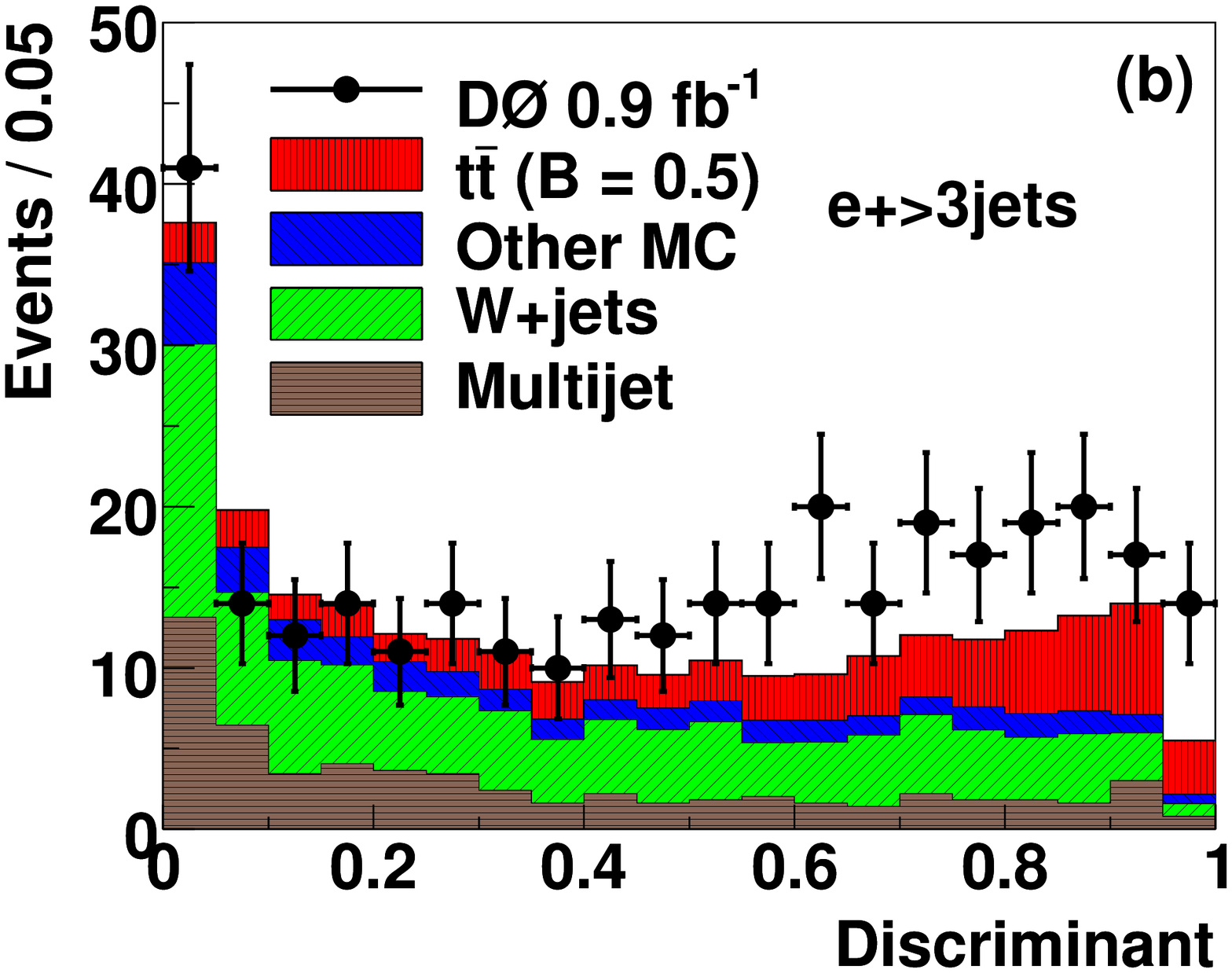} 
\vspace*{-0.2cm}  
\caption{Distributions in the discriminant  $\mathcal{D}$ 
for \mHp~=~120~GeV in $e$ + $> 3$ jets, for (a) $B = 0$ (SM), 
and (b) $B = 0.5$.}
\label{fig:likelihood}
\vspace*{-0.5cm}  
\end{figure}

To differentiate between $\ttbar$ and background, 
we define a multivariate discriminant
\begin{equation}
\label{eq:D}
\mathcal{D}(\mathbf{x}) =
\frac{p(\mathbf{x}|\mathcal{S})}{p(\mathbf{x}|\mathcal{S}) 
+ p(\mathbf{x}|\mathcal{B})},
\end{equation}
where  $p$ is the probability density for a set of observed variables 
$\mathbf{x}$, given the signal ($\mathcal{S}$) 
or background ($\mathcal{B}$) class of events.
The signal comprises $\ttbar$ events which include $H^{\pm}$ decays of 
the top quark and hence depends on the values of $B$ and $m_{H^{\pm}}$. 
For the construction of the discriminant, we define the signal for only one 
value of $B$ that corresponds to the SM scenario ($B = 0$). The background is 
defined by all non-$\ttbar$ events.  The variables for the different final states are listed 
in Table~\ref{tab:var}. The normalization for the $W$+jets template 
is obtained from the low-$\mathcal{D}$ region ($\mathcal{D}<0.45$),  
which is background dominated, by setting the sum of all 
backgrounds and signal in this region to the corresponding observed 
number of events. We recompute the normalization for each value of $B$ and 
$m_{H^{\pm}}$. The number of predicted (and observed) events for the 
full range of $\mathcal{D}$ appears in Table~\ref{tab:yields} for $B=0$. 
The corresponding distributions are shown for \mHp~=~120~GeV in 
Fig.~\ref{fig:likelihood} for $e~+ > 3$ jets, for $B=0$ and 
$B=0.5$, in (a) and (b), respectively. We see that the data agree well 
with the SM predictions. Similar agreement is seen in all  
other channels. Hence, we proceed to set upper limits on the non-SM  
branching fraction $B(\tHb)$.
\begin{table}[!h!tbp]  
\begin{center}  
\caption{Variables used to define the discriminant $\mathcal{D}$. 
$\Delta R=\sqrt{(\Delta\phi)^2+(\Delta\eta)^2}$ and $i$ 
indexes the list of ${N_j}$ jets ordered in decreasing $p_T$.} 
\begin{tabular}{ll}
\hline\hline
Variable & Channel \\ \hline
$\sum_{i=3}^{N_j} p_T(i)$ & all \\ 
$\sum_{i=1}^{N_j} p_T(i)/\sum_{i=1}^{N_j} p_z(i)$ & $e$ + 3 jets, $e~+ > 3$ jets\\ 
$\sum_{i=1}^{N_j} p_T(i)+p_T(e)+\MET$ & $e$ + 3 jets, $e~+ > 3$ jets\\
$\Delta R(\ell, \mathrm{jet1})$ & all\\
$\Delta R(\mathrm{jet1}, \mathrm{jet2})$ & $e~+ > 3$ jets, $\mu~+ > 3$ jets\\
$\Delta\phi(\ell,~$\MET$)$ & $\mu$ + 3 jets, $\mu~+ > 3$ jets\\
$\Delta\phi(\mathrm{jet1},~$\MET$)$ & $e$ + 3 jets, $\mu$ + 3 jets\\
Sphericity {$\mathcal S$}~\cite{topovars} &  all but $\mu$ + 3 jets\\
Aplanarity {$\mathcal A$}~\cite{topovars} &  all but $\mu$ + 3 jets\\
\hline\hline
\end{tabular}
\label{tab:var}
\end{center} 
\vspace*{-0.5cm} 
\end{table} 
%
\begin{table*}[!h!tbp]
\begin{center}
\caption{Event yields after all selections for channels separated 
by lepton flavor and jet multiplicity. We assume $B(\tHb)~=~0$  
so that $\ttbar$ includes only SM decays of the top quarks. The ``other 
MC'' comprises single top quark, diboson, and $Z$+jets events. 
(The uncertainty on the total SM prediction includes correlations 
across samples.)}
\begin{tabular}{lr@{$\,\pm \,$}lr@{$\,\pm \,$}lr@{$\,\pm \,$}lr@{$\,\pm \,$}l}\hline\hline
Source      & 
\multicolumn{2}{c}{$e$ + 3 jets} &
\multicolumn{2}{c}{$\mu$ + 3 jets} &
\multicolumn{2}{c}{$e$ + $> 3$ jets} &
\multicolumn{2}{c}{$\mu$ + $> 3$ jets} \\ \hline
Signal ($\ttbar$) & 148.8 & 20.0 & 108.2 & 14.7  & 130.4 & 19.4 & 105.6 & 15.4\\ 
$W$+jets        & 535.4 & 47.9 & 572.4 & 34.7  & 79.2  & 17.3  & 152.0 & 16.5 \\
Other MC        & 102.5 & 14.6 & 106.7 & 15.3  & 33.1  &  4.8 &  35.0 &  5.3 \\
Multijets       & 194.2 &  30.5 &  33.5 &  13.9  & 60.2 &   10.1 &  10.4 &  5.7 \\ \hline
Total SM prediction & 980.9 & 25.8 & 820.8 & 27.6 & 302.9  & 13.1 & 303.0 & 15.6 \\
Observed & \multicolumn{2}{c}{948} & 
\multicolumn{2}{c}{812} & 
\multicolumn{2}{c}{320}  & 
\multicolumn{2}{c}{306} \\\hline\hline
\end{tabular}
\label{tab:yields}
\end{center}
\vspace*{-0.5cm}
\end{table*}
%
%
%

\begin{figure} [h] 
\includegraphics[width=0.45\textwidth]   
{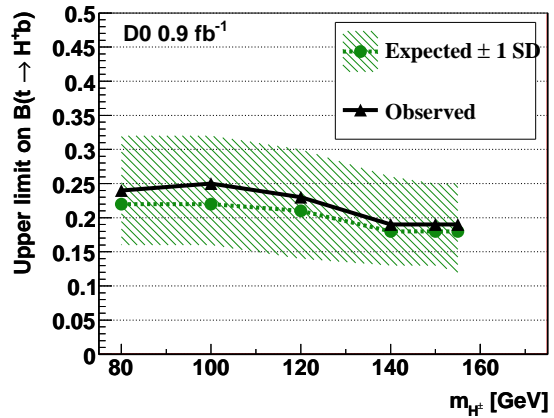}
\vspace*{-0.4cm} 
\caption{The $95\%$ C.L. limits on $B(\tHb)$ for  different values of 
$m_{H^{\pm}}$. }
\label{fig:limitMHvsB}
\vspace*{-0.5cm}   
\end{figure}

We use a modified frequentist approach~\cite{CLS} to set 
limits at the $95\%$ C.L. in the high-$\mathcal{D}$ region 
since it is $\ttbar$ dominated. Sources
of uncertainty on the predicted  yields are 
included with correlations across samples and channels. 
Their estimated values are provided in Table~\ref{tab:sys}. Note that 
all these uncertainties are applied to the $W$+jets normalization 
assuming full anticorrelation because of the manner in which the 
$W$+jets normalization is derived as explained above. The dominant 
sources of uncertainties are from the integrated luminosity, the jet energy 
calibration, and the $\ttbar$ cross section. The uncertainties from the normalization 
of multijet, single top quark, diboson, and $Z$+jets events have a less 
pronounced effect on the $B$ limits because of the smaller contribution 
of these samples in the high discriminant region. We consider the 
distribution in $\mathcal{D}$ above 0.55 in the 
$>3$ jets channels and 0.6 in the $3$ jets channels, a choice determined by 
maximizing the sensitivity of the analysis in MC simulation. The sensitivity 
is defined as the median of limits obtained from an ensemble of 
background plus SM $\ttbar$ ($B = 0$) pseudo experiments in each channel. 
We call these the expected limits and show them by the dashed curve in 
Fig.~\ref{fig:limitMHvsB} along with their $\pm 1$~standard-deviation (SD) 
intervals by the cross-hatched region. The observed limits, using 
D0 data, are shown by the solid curve in Fig.~\ref{fig:limitMHvsB}. 

\begin{table}[!h!tbp]  
\begin{center}  
\caption{Uncertainties (equivalent to $\pm 1$ SD) 
from different components affecting the predicted yields. 
``Other MC'' comprises single top quark, diboson,  and 
$Z$+jets events.} 
\begin{tabular}{lc}
\hline \hline
Component & Uncertainty [$\%$] \\\hline
Integrated luminosity   & 6.1 \\
Primary vertex modeling & 2.2 \\
Trigger efficiency      & 0.5--2.8 \\
Lepton identification   & 2.2--2.6 \\
Jet energy calibration  & 5.0 \\
Jet identification      & 2.0--2.4 \\
Jet energy resolution   & 0.1--1.8 \\
Multijets normalization & 15.7--54.8 \\
Other MC normalization  & 11.0--12.0 \\
$\sigma(\ttbar)$ & 7.4--9.6 \\
$m_t$ & 2.1 \\
MC statistics           & 0.9--25.0 \\\hline\hline
\end{tabular}
\label{tab:sys}
\end{center} 
\vspace*{-0.4cm} 
\end{table}
%

\begin{figure} [htb]
\vspace*{-0.4cm}  
\includegraphics[width=0.45\textwidth]{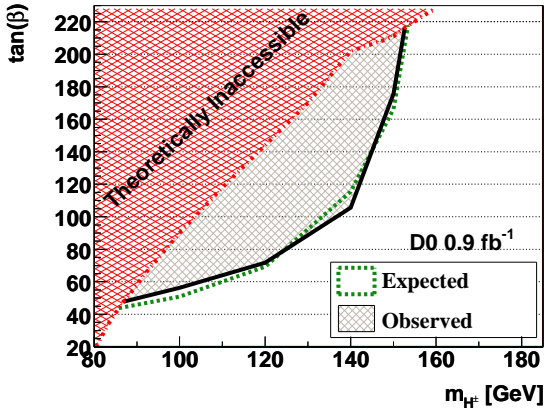}
\vspace*{-0.4cm} 
\caption{The MSSM exclusion regions for the $m^{\max}_h$ scenario.}
\label{fig:mhtanb}
\vspace*{-0.2cm}  
\end{figure}

The upper limit on $B(\tHb)$ can be used to exclude
regions of the [$\tan\beta$, $m_{H^{\pm}}$] parameter space in the context
of the MSSM. Since the MSSM has several free parameters, we
select them according to the $m_h^{\max}$ scenario
described in Ref.~\cite{mhmax}.
This provides the maximum range in the mass of the
lightest neutral Higgs boson as a function of
$\tan\beta$. The exclusion bounds are calculated using 
{\sc FeynHiggs}~\cite{feynhiggs}, which includes the two-loop QCD and
MSSM corrections. 
Figure~\ref{fig:mhtanb} shows the expected and observed excluded 
regions, and the theoretically inaccessible region defined as the boundary 
where certain Higgs parameters acquire unphysical values.

In summary, we have analyzed $0.90 \pm 0.05 ~\rm fb^{-1}$ of 
lepton+jets data at D0, and found no evidence for top quark 
decays to charged Higgs bosons. Hence we set upper limits 
at the $95\%$~C.L. on $B(\tHb)$ ranging from 0.24 for $\mH=80$~GeV 
to 0.19 for $\mH=155$~GeV.

%
We thank the staffs at Fermilab and collaborating institutions, 
and acknowledge support from the 
DOE and NSF (USA);
CEA and CNRS/IN2P3 (France);
FASI, Rosatom and RFBR (Russia);
CNPq, FAPERJ, FAPESP and FUNDUNESP (Brazil);
DAE and DST (India);
Colciencias (Colombia);
CONACyT (Mexico);
KRF and KOSEF (Korea);
CONICET and UBACyT (Argentina);
FOM (The Netherlands);
STFC and the Royal Society (United Kingdom);
MSMT and GACR (Czech Republic);
CRC Program, CFI, NSERC and WestGrid Project (Canada);
BMBF and DFG (Germany);
SFI (Ireland);
The Swedish Research Council (Sweden);
CAS and CNSF (China);
and the
Alexander von Humboldt Foundation (Germany).
%

%



\begin{thebibliography}{99}
\bibitem{d0system}
The D0 coordinate system has the positive 
$z$-axis along the proton beam-line, and $z = 0$ at the center 
of the detector. The polar and azimuthal angles are denoted as 
$\theta$ and $\phi$, respectively. The pseudorapidity is defined 
as $\eta = - \ln(\tan \frac{\theta}{2})$.
\bibitem{topovars}
A momentum tensor is defined as
${\mathcal M}_{ij} = \Sigma_o p^o_ip^o_j/\Sigma_o |\vec{p^o}|^2$,  
where $\vec{p^o}$ is the momentum of a reconstructed object 
$o$, and $i$ and $j$ are Cartesian coordinates. The sum over 
$o$ includes up to four jets of leading $p_T$ and the lepton. 
The sphericity in an event is defined as 
${\mathcal S} = \frac{3}{2}(\lambda_{2} + \lambda_{3})$, 
where $\lambda_{2}$ and $\lambda_{3}$ are the smallest two 
eigenvalues of the normalized momentum tensor.
The aplanarity is defined as ${\mathcal 
A}=\frac{3}{2}\lambda_{3}$.
\end{thebibliography}
\end{document}